\documentclass[oneside,a4paper,11pt,shownumbers]{article}
\usepackage{latexsym}
\usepackage{euscript}
\usepackage{epsfig,amsmath,amssymb}

\newcommand{\e}{{\rm e}}
\def\be{\begin{equation}}
\def\ee{\end{equation}}
\def\bea{\begin{eqnarray}}
\def\eea{\end{eqnarray}}

\long\def\symbolfootnote[#1]#2{\begingroup%
\def\thefootnote{\fnsymbol{footnote}}\footnote[#1]{#2}\endgroup} 

 \large\normalsize

\begin{document}

\title{\LARGE \bf Cylindrically-Symmetric Solutions in Conformal Gravity}
  \author{
  \large  Y. Brihaye$^a$ $\:${\em and}$\:$
  Y. Verbin$^b$ \thanks{Electronic addresses: brihaye@umh.ac.be; verbin@oumail.openu.ac.il } }
 \date{ }
   \maketitle
    \centerline{$^a$ \em Physique Th\'eorique et Math\'ematiques, Universit\'e de Mons,}
   \centerline{\em Place du Parc, B-7000  Mons, Belgique}
     \vskip 0.4cm
   \centerline{$^b$ \em Department of Natural Sciences, The Open University
   of Israel,}
   \centerline{\em Raanana 43107, Israel}
   
\maketitle
\thispagestyle{empty}   

\begin{abstract}
Cylindrically-symmetric solutions in Conformal Gravity are investigated and several new solutions are presented
and discussed.
Among them, a family of vacuum solutions, generalizations of the Melvin solution and cosmic strings of the 
Abelian Higgs model. The Melvin-like solutions have finite energy per unit length, while the string-like solutions 
do not. \\

\noindent{PACS numbers: 04.50.-h, 11.27.+d}

\end{abstract}

\maketitle
\medskip \medskip

\section{Introduction}\label{Introduction}
\setcounter{equation}{0}

Conformal Gravity (CG) \cite{Mannheim2006} was proposed as a possible alternative to Einstein gravity (``GR''), which may 
supply the proper framework for a solution to some of the most annoying problems of theoretical physics like those of the 
cosmological constant, the dark matter and the dark energy.

The gravitational field  in CG is still minimally coupled to matter, but the dynamical basis is different: it is obtained by
 replacing the Einstein-Hilbert action with the Weyl action based on the Weyl (or {\it conformal}) tensor 
 $C_{\kappa\lambda\mu\nu}$ 
defined as the totally traceless part of the Riemann tensor:
\begin{eqnarray}
C_{\kappa\lambda\mu\nu}=R_{\kappa\lambda\mu\nu}-
\frac{1}{2}(g_{\kappa\mu}R_{\lambda\nu}-g_{\kappa\nu}R_{\lambda\mu}+
g_{\lambda\nu}R_{\kappa\mu}-g_{\lambda\mu}R_{\kappa\nu})+
\frac{R}{6}(g_{\kappa\mu}g_{\lambda\nu}-g_{\kappa\nu}g_{\lambda\mu})
\label{WeylTensor},
\end{eqnarray}
so the gravitational Lagrangian is
\begin{equation}
{\cal L}_{g}= -\frac{1}{2\alpha}C_{\kappa\lambda\mu\nu}C^{\kappa\lambda\mu\nu} 
\label{GravL}
\end{equation}
where $\alpha$ is a dimensionless parameter. The gravitational field equations take the following form:
\begin{equation}
W_{\mu\nu} =  \frac{\alpha}{2} T_{\mu\nu} 
\label{GravFieldEq}
\end{equation}
where $T_{\mu\nu}$ is the energy-momentum tensor and $W_{\mu\nu} $ is
the Bach tensor given by:
\begin{eqnarray}
W_{\mu\nu}=\frac{1}{3}\nabla_\mu\nabla_\nu R-\nabla_\lambda\nabla^\lambda R_{\mu\nu}
+\frac{1}{6} (R^2+\nabla_\lambda\nabla^\lambda R-3R_{\kappa\lambda}R^{\kappa\lambda})g_{\mu\nu}+
2R^{\kappa\lambda}R_{\mu\kappa\nu\lambda}-\frac{2}{3}RR_{\mu\nu}
\label{BachTensor}.
\end{eqnarray}

It was suggested (see \cite{Mannheim2006} and references therein) that while CG agrees with Newtonian gravity in 
Solar System scales, it further produces a linearly growing potential that could explain galactic rotation curves 
without invoking dark matter. It was further argued that accelerating cosmological solutions of CG describe naturally the 
accelerated expansion of the universe thus removing the need for dark energy.

On the other hand, CG has been criticized from several aspects both phenomenological and formal. 
Arguments in favor of the need of dark matter come from observations of the unusual object called
``bullet cluster'' \cite{CloweEtAl2006,BradacEtAl2006} whose dynamics seems very difficult to understand without
assuming a weekly interacting dark component.

More specifically, several authors claim that predictions in the weak field limit of CG disagree 
with solar system observations \cite{Flanagan}, yield wrong light deflection \cite{EderyPar1998} and
that the exterior solutions cannot be matched to any source with a ``reasonable'' 
mass distribution \cite{PerlickXu}. Other authors find evidence for tachyons or ghosts 
\cite{BarabashSht1999} or raise the fact that only  null geodesics are physically meaningful in this 
theory since the ``standard'' point particle Lagrangian is not conformally-invariant \cite{WoodMoreau}.

Counter arguments to some of these objections were also published \cite{EderyEtAl2001,Mannheim2007}, and the matter is, to 
our view, still waiting for a consensus. 

A somewhat different approach to the subject, was suggested recently \cite{BrihayeVerbinSph} and solves naturally some of 
the above-mentioned problems, like that of the meaning of time-like geodesics. It is based on extending CG into a scalar-tensor 
theory by introducing an additional real scalar field, so the conformally-invariant point particle Lagrangian will be 
\begin{equation}
L_{pp}=-S\sqrt{g_{\mu\nu}\dot{x}^\mu \dot{x}^\nu} 
\label{L-PointP}
\end{equation}
where $S$ is a real scalar field with the usual conformal transformation laws.

It is therefore very much required to investigate further the predictions and consequences of CG in its purely tensorial
formulation as well as in its scalar tensor extension  as much as possible. 

Following our previous studies of spherically-symmetric solutions \cite{BrihayeVerbinSph} including the non-Abelian 
case \cite{BrihayeVerbinYM} in CG, we move now to cylindrically-symmetric solutions and especially to cosmic strings and we 
start an investigation of their properties in this context. The present report may be considered therefore as a third step of
this program.

Cosmic strings in CG will be the subject of the last section (Sect. \ref{SecStrings}) of this paper. After  a short
general discussion of the field equations of the self-gravitating Abelian Higgs model in the  cylindrically symmetric case
 (Sect. \ref{SecAbHiggs}), we  will start (Sect. \ref{SecCylSymVS}) with vacuum solutions of CG  and line source solutions. 
Then we will discuss purely magnetic solutions (Sect. \ref{SecMelvin}), i.e. the analogs of the Melvin solution 
of ordinary GR and finally we will get to cosmic strings of the Abelian Higgs model. We will see that unlike their GR 
counterparts, the Abelian Higgs string-like solutions in CG although localized, do not have finite energy per unit length.

Since a large part of our discussion will be based on the Abelian Higgs model, we end this section with a few words about its 
coupling to CG.

Among all the higher order gravitational theories, CG is unique in the sense it is based on an additional 
symmetry principle. The conformal symmetry imposes severe limitations on the allowed matter Lagrangian, but the Abelian Higgs 
model is essentially 
still consistent with the conformal symmetry provided the scalar field ``mass term'' is replaced with the appropriate 
``conformal 
coupling'' term which introduces a coupling to the Ricci scalar 
$R$. The matter Lagrangian is therefore
\begin{equation}
{\cal L}_{m}= \frac{1}{2}(D_\mu \Phi)^*(D^\mu
\Phi)-\frac{1}{12}R|\Phi|^2 -\frac{\lambda}{4}|\Phi|^4 
-\frac{1}{4}F_{\mu\nu}F^{\mu\nu} \label{matterL},
\end{equation}
and the resulting field equations are
\begin{eqnarray}
D_\mu D^\mu \Phi + \lambda |\Phi|^2 \Phi + \frac{R}{6} \Phi &=& 0 \\
\nabla_{\mu } F^{\mu\nu} =-\frac{ie}{2}[\Phi^*(D^\nu \Phi)-\Phi (D^\nu \Phi)^*]&=& J^{\nu }\nonumber
\label{FieldEqsScalarVector}.
\end{eqnarray}

The energy-momentum tensor $T_{\mu\nu}$ is given by:
\begin{equation}
{T}_{\mu\nu} = { T}_{\mu\nu}^{(minimal)}+{1\over 6}
\left (g_{\mu \nu} \nabla ^{\lambda }\nabla_
{\lambda } |\Phi|^2 - \nabla _{\mu }\nabla_{\nu } |\Phi|^2  - 
{G}_{\mu \nu} |\Phi|^2 \right)
\label{confTmn}
\end{equation}
${T}_{\mu\nu}^{(minimal)}$ being the ordinary (``minimal'') energy-momentum tensor and ${G}_{\mu \nu}$ is the Einstein tensor.

We can see already at this stage that in order to obtain string-like solutions with $|\Phi|$ approaching asymptotically a 
constant, the geometry cannot be  asymptotically flat. The most that can be obtained are solutions which have asymptotically 
constant and negative Ricci scalar. Actually, a constant Ricci scalar is not a ``gauge invariant'' concept in CG; neither
is a constant scalar field. Thus, it is only a matter of convenience which can be obtained by a proper gauge choice.
Moreover, we would like the gauge choice to be consistent with symmetric vacuum solutions. Thus, we will choose a gauge which
allows to impose the condition
\begin{equation}
R^{\mu}_{\nu}\rightarrow -\frac{\kappa}{4}\delta^{\mu}_{\nu} \;\;\;\; \text{for} \;\; r\rightarrow \infty
 \label{AsymptAdS}
\end{equation}
where $\kappa$ is a positive parameter. We will also limit our solutions to those exhibiting boost symmetry in one direction 
(say, $z$), as for ordinary cosmic strings. These restrictions will simplify considerably the very cumbersome expressions 
of the components of the Bach tensor and will enable a clear physical picture.

The scalar-tensor extension of CG is done along these lines. The gravitational Lagrangian (\ref{GravL}) will be modified to
\begin{equation}
{\cal L}_{g}= \frac{1}{\alpha}\left(-\frac{1}{2}C_{\kappa\lambda\mu\nu}C^{\kappa\lambda\mu\nu}
+\frac{1}{2}\nabla _{\lambda }S\nabla^{\lambda } S-\frac{1}{12}RS^2 -\frac{\nu}{4}S^4  \right)  
\label{STGravL}
\end{equation}
where $S$ is a real scalar field with the usual conformal transformation laws and $\nu$ is a possible self-coupling parameter.
This should go together with postulating the conformally-invariant point particle Lagrangian, (\ref{L-PointP}). Point particles
couples in this picture to the ``physical metric'' $S^2 g_{\mu\nu}$.

However, when the Higgs field is present, we may identify $S$ with $|\Phi|$ so having a more economic model where the Higgs field
is responsible for the mass of point particles also in this classical framework.

\section{Cylindrically-Symmetric Solutions} \label{SecAbHiggs}
\setcounter{equation}{0}

The general cylindrically symmetric line-element has the form:
\begin{equation}
ds^2= B^2 (r)dt^2 - M^2 (r)dr^2 - L^2 (r)d\varphi^2 - K^2 (r)dz^2
\label{lineel}
\end{equation}
and the general expression for the Ricci tensor turns out to be
\begin{eqnarray}
R^0_0 = -\frac{1}{BLKM}\left(\frac{KLB'}{M}\right)'
\,\, &,&\,\,    R^r_r = -\frac{1}{M^2}\left(\frac{B''}{B}+\frac{L''}{L}+\frac{K''}{K}-\frac{M'}{M}\frac{(KLB)'}{KLB}\right)  
\nonumber \\ 
R^\varphi_\varphi = -\frac{1}{BLKM}\left(\frac{KBL'}{M}\right)' 
\,\, &,&\,\, R^z_z = -\frac{1}{BLKM}\left(\frac{BLK'}{M}\right)' 
\label{CylRicciT}
\end{eqnarray}
with the Ricci scalar
\begin{equation}
R =\frac{2}{M^2}\left[-\frac{B''}{B}-\frac{L''}{L}-\frac{K''}{K}
-\frac{B'}{B}\frac{L'}{L}-\frac{L'}{L}\frac{K'}{K}-\frac{K'}{K}\frac{B'}{B}
+\frac{M'}{M}\frac{(KLB)'}{KLB}\right]
\label{RicciS}
\end{equation}

Since all the solutions that we will consider in this work can be thought technically as special cases of the solutions of
the Abelian Higgs model coupled to CG, we discuss now the field equations of this system. 

The general cylindrically symmetric matter fields are parametrized by
\begin{equation}
\Phi=f(r)e^{im\varphi} \,\,\,; \,\,\, eA_{\mu}dx^{\mu}=A(r)d\varphi
\label{CylSymmMatter}.
\end{equation}

Using the above parametrizations, the field equations for the scalar and vector fields get the following simple form:
\begin{eqnarray}
f''+\left(\frac{2B'}{B}+\frac{L'}{L}\right)f'-\left[\left(\frac{m-A}{L}\right)^2+ \frac{R}{6}+
\lambda f^2 \right]f =0  \label{ScFEqsCylSymm} \\
A''+\left(\frac{2B'}{B}-\frac{L'}{L}\right)A'+e(m-A) f^2=0
\label{VecFEqsCylSymm}.
\end{eqnarray}
The gravitational part of the field equations is less simple due to the complicated form of the Bach tensor. 
First we write explicitly the non-vanishing components of $T^{\mu}_{\nu}$ and in order to simplify matters we 
define the following partial energy densities:
\begin{eqnarray}
\varepsilon_s = \frac{\left(f'\right)^2}{2} \ \ ,\hspace{8 mm}
\varepsilon_v = {1\over 2}\left(\frac{A'}{L}\right)^2 ,\hspace{8 mm} 
\varepsilon_{sv} = {1\over 2}\left(\frac{m-A}{L}\right)^2 f^2 ,\hspace{8 mm}
u = \frac{\lambda}{4} f^4 
\label{densities}.
\end{eqnarray}
In terms of these the components of $T^{\mu}_{\nu}$ will be
\begin{eqnarray} 
     T^{0}_{0} \ = \ T^{z}_{z} &=& \varepsilon _v + {1\over 3}(\varepsilon _s + \varepsilon _{sv} - u )
     -\frac{f^2}{6} \left( R^0_0 - \frac{R}{6} \right)+\frac{ff'}{3}\frac{B'}{B}  \nonumber \\
     T^{r}_{r} &=& -\varepsilon _v -\varepsilon _s + \varepsilon _{sv} + u 
-\frac{f^2}{6} \left( R^{r}_{r} - \frac{R}{2} \right)-\frac{ff'}{3}\left(\frac{2B'}{B}+\frac{L'}{L}\right) \nonumber \\
 T^{\varphi}_{\varphi} &=& -\varepsilon _v + {1\over 3}(\varepsilon _s -5\varepsilon _{sv} - u ) 
-\frac{f^2}{6} \left( R^{\varphi}_{\varphi} - \frac{R}{6} \right)+\frac{ff'}{3}\frac{L'}{L}
 \label{TmnCyl}
 \end{eqnarray}
where the Ricci tensor components are given in Eq. (\ref{CylRicciT}).
As an easy check one can verify that indeed the sum of these components vanishes as it should. Note that unlike in 
GR, this does not force a vanishing Ricci scalar since the gravitational field equations are 
given by Eq. (\ref{GravFieldEq}). In the present case they reduce to two components  of (\ref{GravFieldEq}) which we
may choose to be the $tt$ and $rr$ components: 
\begin{eqnarray} \nonumber
 W^{0}_{0} \  &=& \frac{\alpha}{2}\left[\varepsilon _v + {1\over 3}(\varepsilon _s + \varepsilon _{sv} - u )
     -\frac{f^2}{6} \left( R^0_0 - \frac{R}{6} \right)+\frac{ff'}{3}\frac{B'}{B}\right]  \nonumber \\
     W^{r}_{r} &=& \frac{\alpha}{2}\left[-\varepsilon _v -\varepsilon _s + \varepsilon _{sv} + u 
-\frac{f^2}{6} \left( R^{r}_{r} - \frac{R}{2} \right)-\frac{ff'}{3}\left(\frac{2B'}{B}+\frac{L'}{L}\right)\right]
\label{GravFieldEqCYL0}.
 \end{eqnarray}
We still refrain from writing explicitly the components of $W^\mu_\nu$ anticipating further simplifications.

 In order to find solutions for this system, we have to fix the arbitrariness of the radial 
coordinate and the arbitrary rescaling of the metric due to the conformal symmetry. This, together with the boost
symmetry in the $tz$ plane leaves one independent metric component. So only one of the gravitational field 
equations has to be solved. We chose to solve the lower order $rr$ equation of (\ref{GravFieldEqCYL0}), and used the higher 
order $tt$ equation as a consistency check.

The choice of the conformal rescaling further has to be compatible with the condition 
$R_{\mu}^{\nu} \propto \delta_{\mu}^{\nu}$ asymptotically (see (\ref{AsymptAdS})). This is not obvious even in
vacuum: for instance,  in the cylindrical version of the ``Mannheim gauge''\cite{Mannheim2006}  
$
 M(r)=1/B(r) \; , L(r) = r 
$
with $K(r) = B(r)$, the vacuum equations are inconsistent; in the gauge 
$
M(r)=1, K(r)=1 \ , 
$
the vacuum equations are consistent but the  solutions are such that the diagonal components 
of Ricci tensor depend linearly on $r$ for $r\to\infty$. 
  
One convenient ansatz which solves these problems is
\be
\label{gauge}
      M = 1 \ \ , \ \ L=\frac{dB}{dr} \ \ , \ \ K(r) = B(r)
\ee
where we also resort to dimensionless coordinates. 

Moreover, all the thin string (line source) solutions of GR with a negative cosmological constant \cite{linet} which solve 
for $r>0$
\be
R^{\mu}_{\nu}= -\frac{\kappa}{4}\delta^{\mu}_{\nu}
\label{AdSEq} 
\ee
 satisfy also the CG vacuum equations $W_{\mu\nu}=0$.
Among those solutions is also found the (four dimensional) AdS soliton \cite{HorMy1999} (see also \cite{bbh,Bonnor,tekin}) 
which is a cylindrically-symmetric regular solution of the same equation 
(\ref{AdSEq}) and is therefore distinct from AdS (anti-de Sitter) space.

In this gauge the components of the Bach tensor are 
\begin{eqnarray} \nonumber
W_0^0  =W_z^z  =
\frac{B^{(5)}}{3 B'}+\frac{2 B^{(4)}}{3 B}-
\frac{B^{(4)} B''}{3 B'^2}+\frac{B''' B''^2}{3 B'^3}+\frac{2 B''' B''}{3 B B'}-\frac{2 B''' B'}{3 B^2}\\
-\frac{2 B'''^2}{3 B'^2}-\frac{4 B''^2}{3 B^2}+\frac{4 B'^2 B''}{3 B^3}-
\frac{B'^4}{3 B^4}
\label{Bach00}
\end{eqnarray}

\begin{eqnarray} \nonumber
W_r^r  =\frac{2 B^{(4)}}{3 B}-
   \frac{2 B^{(4)} B''}{3 B'^2}+\frac{2 B''' B''^2}{3 B'^3}-\frac{2 B'''B''}{B B'} +\frac{2 B''' B'}{3B^2}\\
  +\frac{B'''^2}{3B'^2}
   +\frac{4 B''^2}{3 B^2}-\frac{4 B'^2 B''}{3B^3}+
\frac{B'^4}{3B^4}
\label{Bachrr}
\end{eqnarray}
while the fourth one, $W^\varphi_\varphi$ can be obtained immediately from the identity $W^\mu_\mu =0$.

In terms of $B(r)$, the Ricci scalar and tensor take the form   
\be
 R_0^0 = R_z^z = -\left(\frac{B'}{B}\right)^2 - 2\frac{B''}{B}  \  , \ \ 
 R_r^r = R_\varphi^\varphi = -\frac{B'''}{B'} - 2\frac{B''}{B} \  , \ \ 
 R = -2 \left[\left(\frac{B'}{B}\right)^2 + 4 \frac{B''}{B} + \frac{B'''}{B'}\right]
 \label{Ricci-BG}
\ee

Returning now to the general Abelian Higgs system we are therefore left with  the 
 $rr$ component   of equations (\ref{GravFieldEqCYL0}): 
\begin{eqnarray} \nonumber
\frac{2 B^{(4)}}{3 B}-
   \frac{2 B^{(4)} B''}{3 B'^2}+\frac{2 B''' B''^2}{3 B'^3}-\frac{2 B'''B''}{B B'} +\frac{2 B''' B'}{3B^2}
  +\frac{B'''^2}{3B'^2} +\frac{4 B''^2}{3 B^2}-\frac{4 B'^2 B''}{3B^3}+
\frac{B'^4}{3B^4} =  \\
\frac{\alpha}{2}\left[-{1\over 2}\left(\frac{A'}{L}\right)^2 -\frac{\left(f'\right)^2}{2} 
+ {1\over 2}\left(\frac{m-A}{L}\right)^2 f^2  + \frac{\lambda}{4} f^4 
-\frac{f^2}{6} \left( R^{r}_{r} - \frac{R}{2} \right)-\frac{ff'}{3}\left(\frac{2B'}{B}+\frac{L'}{L}\right)\right]
\label{GravFieldEqCYL}
 \end{eqnarray}
which should be solved together with the two equations (\ref{ScFEqsCylSymm})-(\ref{VecFEqsCylSymm}).

\section{Vacuum Solutions} \label{SecCylSymVS}
\setcounter{equation}{0}

We start by considering the vacuum equation $W_{\mu \nu}=0$ associated with the metric (\ref{lineel}) 
in the gauge (\ref{gauge}).
The relevant equation for the function $B(r)$ is just a special case of Eq (\ref{GravFieldEqCYL}), leading to
\begin{eqnarray} 
\frac{2 B^{(4)}}{3 B}-
   \frac{2 B^{(4)} B''}{3 B'^2}+\frac{2 B''' B''^2}{3 B'^3}-\frac{2 B'''B''}{B B'}
   +\frac{2 B''' B'}{3B^2}+
\frac{B'''^2}{3B'^2}
   +\frac{4 B''^2}{3 B^2}-\frac{4 B'^2 B''}{3B^3}+
\frac{B'^4}{3B^4} = 0
\label{equab}.
\end{eqnarray}
Equation (\ref{equab}) has very interesting properties, namely it is autonomous and 
invariant under rescaling of $B$ and of $r$. 

The above-mentioned AdS soliton is just one member of a continuous
family of solutions of (\ref{equab}) (as well as of (\ref{AdSEq})) given by
\be
\label{AdSSolFam}
    B(r) = \left( a\cosh(kr) +b\sinh(kr) \right)^{2/3}.
\ee
The case $k=\sqrt{3/2}$ and $b=0$ corresponds to the AdS soliton which satisfies Eq. (\ref{AdSEq}) with $\kappa=8$. 
The other values of $k$ with $b=0$ correspond to thin string solutions which are the counterparts of the conic 
solutions for $\kappa=0$. The solutions with $a=0$ are analogous to the ``Melvin branch'' of thin strings. 
These have a power law behavior near the string axis with the same powers of $(2/3,\ -1/3,\ 2/3)$ as for $\kappa=0$. 
Actually these exact power law $\kappa=0$ solutions also solve Eq. (\ref{equab}) and may be viewed as a 
limit of (\ref{AdSSolFam}) where $k\rightarrow 0$ and $kb$ remains finite:
\be
\label{AdSpower}
    B(r) = \left(a+br\right)^{2/3}.
\ee
The complexified version of (\ref{AdSSolFam}), namely 
\be
\label{AdSClosed}
    B(r) = \left(a\cos(kr) +b\sin(kr)\right)^{2/3}
\ee
are closed solutions in analogy with the ``inverted cones'' of the $\kappa=0$ case. 

Since the gravitational field equations of CG are of higher order, it is expected that more vacuum solutions exist, i.e.
new vacuum solutions which are special to CG. Indeed the following family also solves Eq. (\ref{equab}) -- but 
satisfies (\ref{AdSEq}) only asymptotically:
\be
\label{CGVacSolFam}
   B(r) = \left(a\cosh(kr) +b\sinh(kr)\right)^{2}.
\ee
Eq.  (\ref{equab}) is solved also by their complexified counterparts, 
\be
\label{AdSClosed2}
    B(r) = \left(a\cos(kr) +b\sin(kr)\right)^{2}
 \ee
as well as by the $k\rightarrow 0$ limit solutions whose Ricci scalar vanishes asymptotically:     
\be
\label{AdSpower2}
    B(r) = \left(a+br\right)^{2}.
\ee

We have again a similar pattern of two families of string-like solutions: an open one and a closed one, 
each of them with the free parameters $a$, $b$ and $k$. The family of open solutions
with $b=0$ contains as before a soliton-like regular solution for the specific value $k=1/\sqrt{2}$. This one
is however a new solution special to CG. We write both solitons explicitly: 
\begin{eqnarray} \nonumber
\label{analytic}
  \kappa = 8 :  \ \ \   B(r) = \cosh^{2/3}\left(\sqrt{3/2} \text{ }r \right) \  &,&  \ R(r) = - 8  \\
  \kappa = 24 :  \ \ \ \ \ \ \ \
    B(r) = \cosh^2\left(r/\sqrt{2}\right)  \  &,& \  R(r) = -12 \left(1 + \tanh^2\left(r/\sqrt{2}\right)\right).    
\end{eqnarray}

\begin{figure}[!t]
\centering
\leavevmode\epsfxsize=10.0cm
\epsfbox{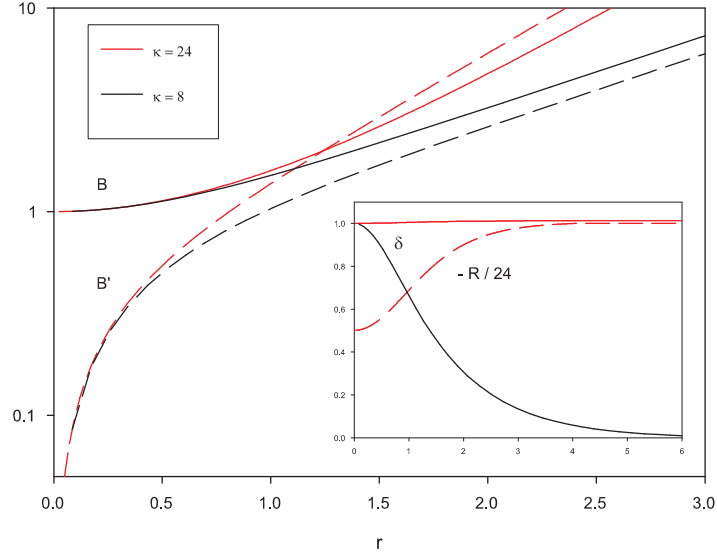}\\
\caption{\label{melvin1_a} \small{
The two analytical regular solutions of the vacuum equations.  Profiles of $B(r),\ L(r)=B'(r)$ and 
of $R(r),\ \delta(r)$. The line $R(r)=-8$ for $\kappa=8$ is not shown.}} 
\end{figure}

It is  likely that equation (\ref{equab}) is integrable by quadrature
but, so far, we were unable to find the general solution analytically. One additional explicit solution that 
we found is $B(r) = e^{\pm kr}$ which is actually conformally flat and thus 
uninteresting for us. In absence of knowledge about the general set of analytical solutions
we investigated the behavior of the solutions of Eq. (\ref{equab}) numerically. 
Exploiting, the different scale invariances of (\ref{equab}), the boundary conditions of regular
solutions can be specified according to
\be
\label{bc}
     B(0) = 1 \ \ , \ \ B'(0) = 0 \ \ , \ \ B''(0) = 1 \ \ , \ \ R(r\to \infty) = - \kappa
\ee 
where $\kappa$ is a positive constant. The two regular solutions (\ref{analytic}) already 
obey the boundary conditions (\ref{bc}). Note the residual scaling symmetry of 
any solution obeying (\ref{bc}) realized by the one-parameter family of solutions 
\be
    B(r) \longrightarrow \hat B(r) = \lambda B(\frac{r}{\sqrt{\lambda}}) \ \ \ , \ \ \ 
    \kappa \longrightarrow \hat \kappa = \frac{\kappa}{\lambda}  \ . \
\ee  

The combination $\delta(r) \equiv (B B'' - (B')^2)/B$ has some relevance in the analysis of the solutions.
In particular the zeros of this function constitute singular points of Eq. (\ref{equab}). The occurrence of such points 
renders the solutions problematic. For $\kappa < 8$, our numerical analysis reveals that the function $\delta(r)$
develops nodes and suggests that no regular solution on $r \in [0,\infty]$ can be constructed.
The boundary conditions lead to $\delta(0)=1$.

The  functions $B(r)$ of (\ref{analytic}) and their derivatives 
$B'(r)=L(r)$  are plotted on Fig. \ref{melvin1_a}. We also added plots of $R(r)$ and $\delta(r)$.

We then attempted to construct numerically the  solutions obeying (\ref{equab})+(\ref{bc}) 
for different values of $\kappa$. It turns out that regular solutions exist only for $\kappa \geq 8$. 
The interpretation of the numerical solutions relies on the understanding of their asymptotic behavior.
The solutions obtained  obey the following asymptotic behavior
\be
\label{asymptotic}
     \log(B(r\to \infty)) =  \gamma_0 + \gamma_1 r + \gamma_2e^{-\beta r} + O(e^{-2 \beta r}) \ \ 
     , \ \      \gamma_1 = \sqrt{\frac{\kappa}{12}} \ \ 
\ee
where $\gamma_0$, $\gamma_2$ and $\beta$ are positive constants.
The relevant component of the Bach tensor is
\be
 W_r^r(r \to \infty) = (3\gamma_1 - \beta) (\gamma_1 - \beta) \frac{\beta^4 \gamma_2^4}{3\gamma_1^1} \e^{-2\beta r}
 \ \ \ , \ \ \ 
\ee
and, for the Ricci scalar, we have 
 \be
    R(r \to \infty) = -12 \gamma_1^2 + 2\frac{\gamma_2 \beta}{\gamma_1}(3\gamma_1 - \beta)(4\gamma_1- \beta) e^{-\beta r}
\ee
Actually, this asymptotic form is general enough to be useful for understanding the solutions where matter is present 
(see next sections).

The numerical results give  $\beta =  \gamma_1$ for the vacuum solutions which is indeed consistent with the vacuum condition
$W_{\mu\nu}=0$. The quantitative values of the other parameters can be extracted from the numerical solutions.
 The value $\gamma_2$ is related to the parameter  $\delta \equiv \delta(\infty)$:
 $\gamma_2 = 12 \frac{\delta}{\kappa} e^{-\gamma_0} $.
   
\begin{figure}[!b]
\centering
\leavevmode\epsfxsize=10.0cm
\epsfbox{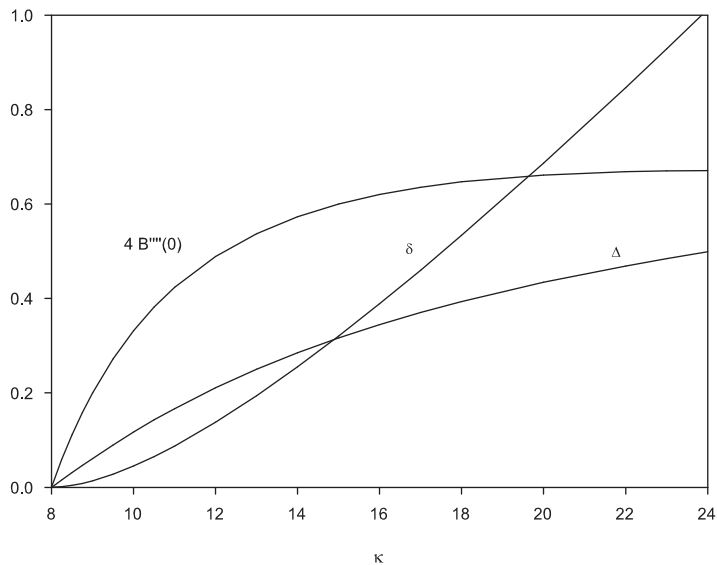}\\
\caption{\label{melvin2} \small{
The value $B''''(0)$, $\delta \equiv \delta(\infty)$ and $\Delta \equiv 1-R(0)/R(\infty)$ as function of $\kappa$.
 Note: $\kappa \geq 8$.}}
\end{figure}

Several parameters characterizing the vacuum solutions are presented as functions of $\kappa$ on Fig. \ref{melvin2};
namely the value $B''''(0)$, the difference $\Delta \equiv R(0)-R(\infty)$ and the parameter 
$\delta$ defined above. In fact, regular solutions exist for $\kappa \geq 8$ but the 
picture is limited  to $\kappa \leq  24$.
We notice that the quantities presented on Fig. \ref{melvin2} all tend to zero in the limit $\kappa \to 8$
corresponding to the AdS soliton. 

\begin{figure}[!t]
\centering
\leavevmode\epsfxsize=10.0cm
\epsfbox{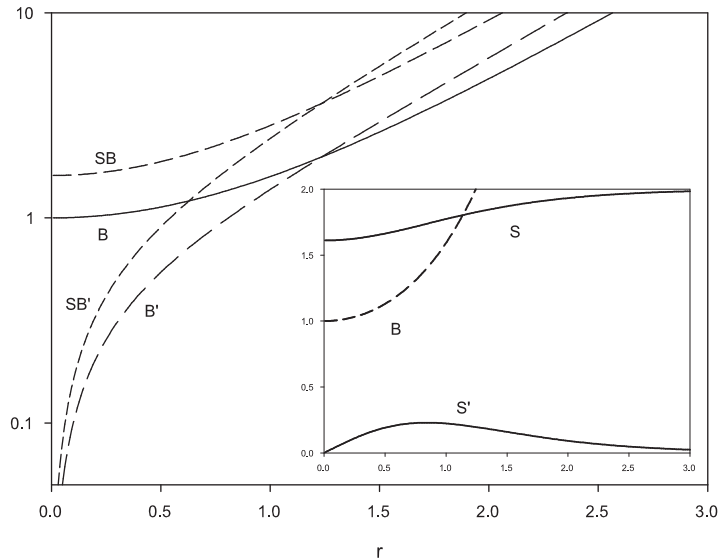}\\
\caption{\label{STVac} \small{A vacuum solution of scalar-tensor CG. Profiles of the scalar field $S(r)$, 
the metric components $B(r)$ and $L(r)=B'(r)$ and those of the ``physical metric'' $S(r)B(r)$  and $S(r)B'(r)$.
$\kappa = 24$ and $\nu =1$.}}
\end{figure}

Finally, we consider the vacuum solutions of the scalar-tensor CG, namely those which are derived
from the Lagrangian (\ref{STGravL}). Technically, they are obtained easily from Eq. (\ref{GravFieldEqCYL}) by 
substituting $A(r)=0$ and $m=0$ and replacing $f(r)$ by $S(r)/\sqrt{\alpha}$  and $\lambda$ by $\alpha\nu$.

We find regular soliton-like solutions by imposing the boundary conditions (\ref{bc}) together with
\be
S'(0)=0   \ \ , \ \ S(r\to \infty) = S_{0}  
\label{S-BC} 
\ee
where $S_{0}$ fixes the asymptotic value of the Ricci scalar by $R(\infty) = -6 \nu S^2_{0}$. Fig. \ref{STVac}
contains the profiles of $S(r)$ and the metric components of a typical solution as well as the ``physical metric'' 
$S^2 g_{\mu\nu}$ which 
is the one which directly couples to point particles - see (\ref{L-PointP}). We notice that since the scalar field $S(r)$
does not change very much, the ``physical metric'' is quite similar to the ``conformal metric''. Moreover,
even the difference with respect to the vacuum metric of the purely tensor theory shown in Fig. \ref{melvin1_a} is not very 
pronounced.

\section{Melvin-like Solutions} \label{SecMelvin}
\setcounter{equation}{0}

\begin{figure}[!b]
\centering
\leavevmode\epsfxsize=10.0cm
\epsfbox{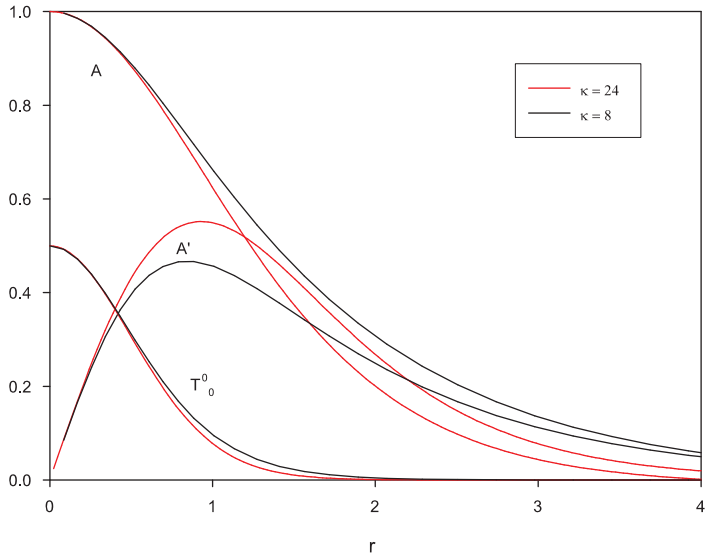}\\
\caption{\label{melvin1_b} \small{
 Profiles of the magnetic potential $A,\ A'$ and of the energy density $T_0^0$
in the background of the two analytic solutions (\ref{analytic}).}}
\end{figure}

Setting $\Phi=0$ in (\ref{matterL}), we obtain the  Weyl-Maxwell Lagrangian
and the corresponding  cylindrically symmetric equations are obtained by setting $f(r)=0$ in
Eqs. (\ref{VecFEqsCylSymm}) and (\ref{GravFieldEqCYL}).
The generalization of the Melvin solution to CG can be looked for.
In fact the Maxwell equation can be integrated directly, leading to $A(r) = Q_m /B(r)$
where $Q_m$ is an integration constant which encodes the magnitude of the magnetic field.
Actually, the $z$-component of the magnetic field, given by ${\cal B}_z(r) = A'_{\varphi}(r)/L(r)$, 
leads to ${\cal B}_z(0)=Q_m$. Also the choice $A\to 0$ as $r \to \infty$ was taken. 
 
The different components of the energy momentum tensor can then be computed in terms of $B(r)$, namely
\be
T_0^0 = -T_r^r  = \varepsilon _v = \frac{{\cal B}_z(0)^2}{2B^4(r)} \ \ .
\ee
The inertial mass per unit length of the conformal Melvin solution can then be computed:
\be
     M_I = \int d^{2}x \sqrt{-g} \  T^0_0 =  2 \pi {\cal B}_z(0)^2 \int_0^{\infty} dr \frac{B'}{2 B^2} =  \pi {\cal B}_z(0)^2
\ee
 
\subsection{Background Solutions: $\alpha=0$}  \label{Sec4.1}
Setting first $\alpha = 0$, leads to solutions in the background of a CG vacuum.
The profiles of the  magnetic functions $A,A'$ corresponding to ${\cal B}_z(0)=1$
are shown in Fig. \ref{melvin1_b} for the  
cases $\kappa=8$ and $\kappa = 24$ 
corresponding to the 
two analytic solutions (\ref{analytic}). The energy density $T_0^0$ is supplemented on the figure.
Similar solutions exist for generic values of $\kappa$,
the profiles of the  magnetic function $A$ depend only weakly on $\kappa$.


\begin{figure}[!t]
\centering
\leavevmode\epsfxsize=8.0cm
\epsfbox{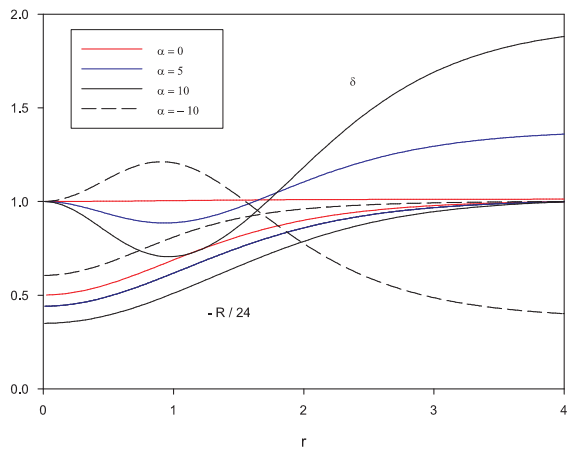}\\
\caption{\label{melvin3} \small{
Melvin-like solutions for $\kappa = 24$: Form of $R(r)$ and $\delta(r)$ for four values of $\alpha$. 
For $\alpha<0$ see Sect. \ref{Sec4.3}.}}
\end{figure}
\begin{figure}[!b]
\centering
\leavevmode\epsfxsize=8.0cm
\epsfbox{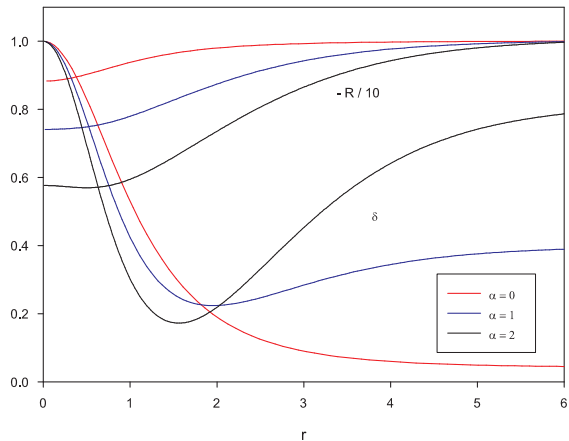}\\
\caption{\label{melvin4} \small{
Melvin-like solutions for $\kappa = 10$: Form of $R(r)$ and $\delta(r)$  for three values of $\alpha$.}}
\end{figure}

\subsection{Gravitating Solutions: $\alpha > 0$} \label{Sec4.2}
Next, we investigate the back-reaction of the electromagnetic field on gravity
by solving the coupled equations (\ref{VecFEqsCylSymm})+(\ref{GravFieldEqCYL}), for $\alpha >0$. 
Using the Maxwell equation to eliminate $A$, we obtain a single equation for $B(r)$ which reads
\be
\label{melvin}
\frac{2 B^{(4)}}{3 B}-
   \frac{2 B^{(4)} B''}{3 B'^2}+\frac{2 B''' B''^2}{3 B'^3}-\frac{2 B'''B''}{B B'} +\frac{2 B''' B'}{3B^2}\\
  +\frac{B'''^2}{3B'^2}
   +\frac{4 B''^2}{3 B^2}-\frac{4 B'^2 B''}{3B^3}+
\frac{B'^4}{3B^4} = -\frac{\alpha {\cal B}_z(0)^2}{4B^4}. 
\ee   
The constant ${\cal B}_z(0)$ can be clearly absorbed here in $\alpha$ so from now on, we set ${\cal B}_z(0)=1$ and study the 
solutions of (\ref{melvin}) for different values of $\alpha$ and $\kappa$.

It should also be  noticed that the function $B(r)$ keeps asymptotically the form (\ref{asymptotic}) because $W_r^r$
in the left hand side  in  Eq. (\ref{melvin}) is asymptotically of order $\exp(-2 \beta r)$ 
with $\beta = \sqrt{\kappa /12}$  while $T_r^r$ in   the right hand side  is asymptotically of order $\exp(-4 \beta r)$. 
The deviation due to the matter field therefore comes out as a next to leading order correction.

In fact, the deviation of the metric and matter functions with respect to the case $\alpha=0$
is not substantial. The deformation of the geometry by the matter fields 
appears more clearly on the Ricci scalar $R(r)$ and on the function
$\delta(r)$ defined in Sect. \ref{SecCylSymVS}.  Fig. \ref{melvin3} shows these functions for
$\kappa = 24$ and for  three positive values of $\alpha$.  
Our results further show that the solutions exist up to a maximal value of $\alpha$,
$\alpha_c \approx 12.1$. 
As another example, for $\kappa=10$, we find $\alpha_c \approx 2.6$. The $r$-dependence of $R(r)$ and $\delta(r)$ 
for $\kappa=10$ and for three values of $\alpha$ is illustrated on Fig.  \ref{melvin4}. 
This feature seems to occur for all values of $\kappa$ and we conclude that for fixed $\kappa$, the solutions exist
up to a maximal value of $\alpha = \alpha_c(\kappa)$ forming a branch of solutions which we denote 'branch I' for 
later convenience.  

Remarkably, the domain of existence of solutions in the $\alpha,\kappa$
plane is determined by
\be
\alpha < \alpha_c(\kappa) = 0.675 \kappa - 4.1
\ee
with a very good accuracy.  
It looks as if, for a fixed $\kappa$, CG cannot support a magnetic
field of arbitrary large strength. In other words,  the value $|{\cal B}_z(0)|$ has to be smaller than a critical value
 (recall absorbing ${\cal B}_z(0)$ in $\alpha$).

\begin{figure}[!b]
\centering
\leavevmode\epsfxsize=10.0cm
 \includegraphics[height=.30\textheight,width=.48\textwidth]{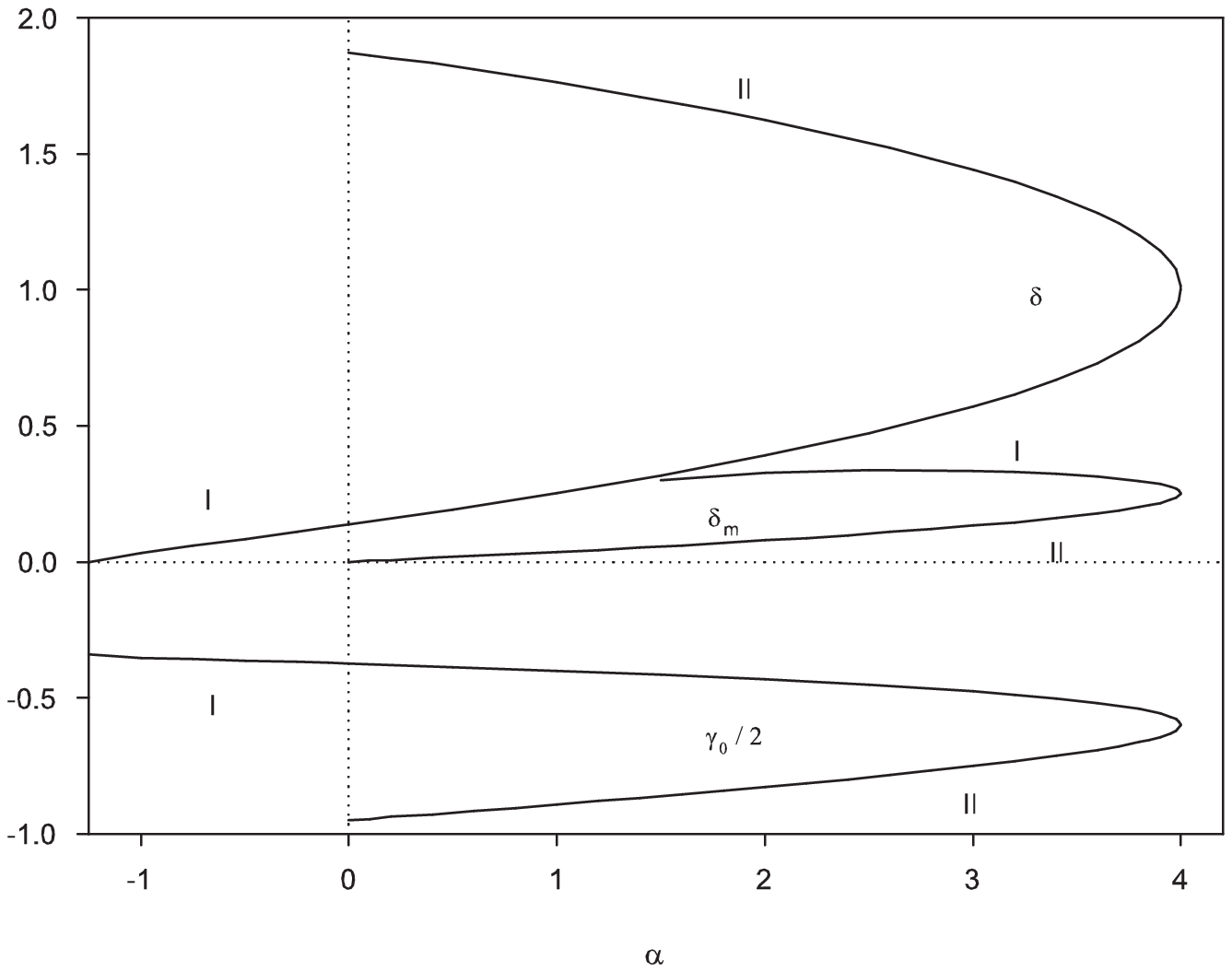}  
 \includegraphics[height=.30\textheight,width=.48\textwidth]{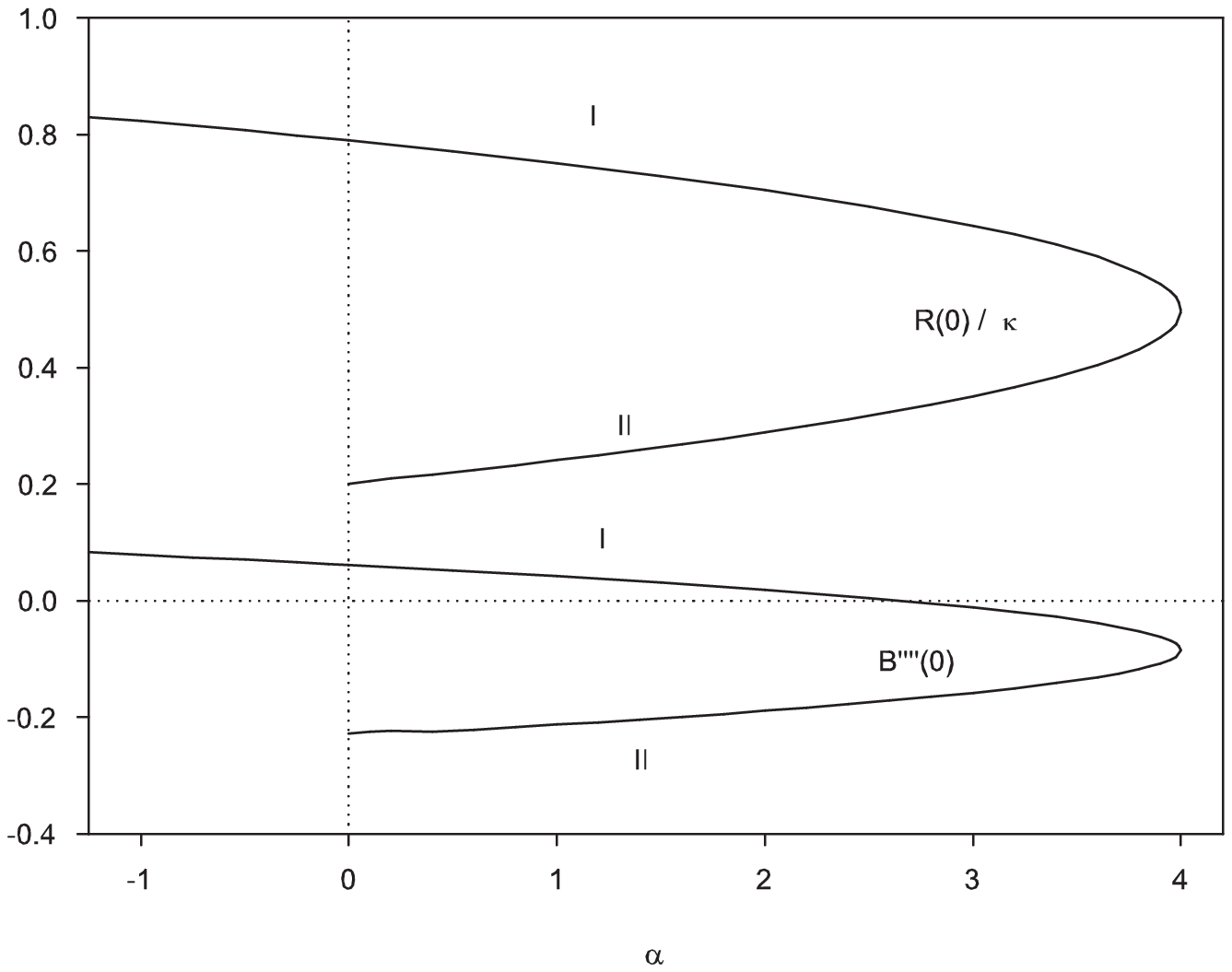}\\
 (a)\hskip 7.5cm (b)\\
 \caption{\label{critical0-1} \small{Melvin-like solutions  for $\kappa=12$:
$\alpha$-dependence of: (a) $\delta \equiv \delta(\infty)$, $\gamma_0$ and $\delta_m$;
(b) $B''''(0)$ and $R(0)/ \kappa$.
The labels I and II distinguish the two branches. For $\alpha<0$ see Sect. \ref{Sec4.3}.}}
\end{figure}  

In order to complete the pattern of the solutions of Eq. (\ref{melvin}), we looked for another family of solutions.
The numerical analysis reveals, indeed,  that  a second branch (to which we refer as 'branch II') 
of self-gravitating  solutions exist for $\alpha \in ]0,\alpha_c]$. In the limit $\alpha \to \alpha_c$
the two branches join, forming a cusp and no solutions exist for $\alpha > \alpha_c$. 

A natural point which raises at this stage is understanding the behavior of the solutions of branch II
for $\alpha \to 0$. It turns out that for sufficiently large values of $\alpha$, the value $B''''(0)$ becomes negative
and the function $\delta(r)$ develops a local minimum, say for $\delta(r_0)= \delta_m$. 
The results further reveal that, in the limit $\alpha \to 0$ the solutions of branch II
are such that $\delta_m \to 0$. As a consequence, the equation develops a singular point in this limit 
and regular solutions no longer exist. 

These phenomena are illustrated in Figs. \ref{critical0-1}a and \ref{critical0-1}b 
where several parameters characterizing the solutions are presented as functions of $\alpha$ for
$\kappa=12$. 

\begin{figure}[!t]
\centering
\leavevmode\epsfxsize=10.0cm
\epsfbox{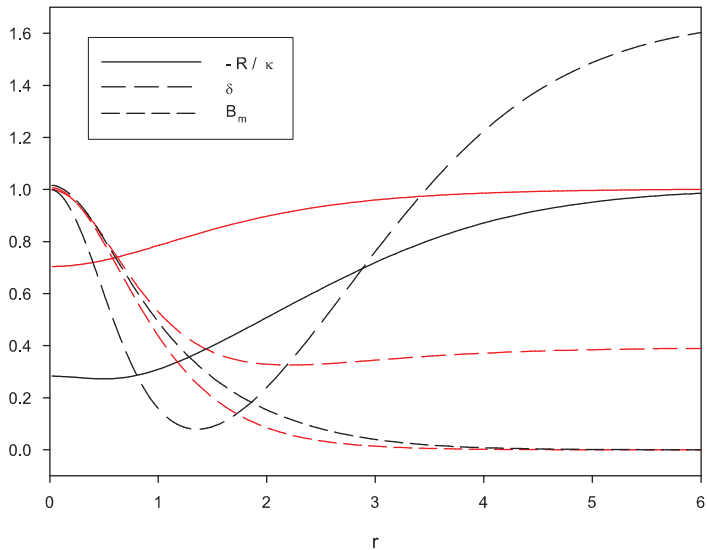}\\
\caption{\label{two_solu} \small{
Comparison of $R(r)$, $\delta(r)$ and of the magnetic field $B_m(r)$ for the two solutions with $\kappa=12$, $\alpha=2$.
The red (resp. black) curves refer to branch I (resp. branch II)}.}
\end{figure}

A comparison between the two solutions corresponding to $\kappa=12$, $\alpha=2$
is  shown on Fig. \ref{two_solu}. The figure reveals that the rather tiny difference occurring for the
matter fields has a significant influence on the geometry;
this appears clearly  on the shape of the Ricci scalar $R(r)$.

\subsection{Gravitating Solutions: $\alpha < 0$} \label{Sec4.3}

The solutions of Sect. \ref{Sec4.1} can be deformed for negative values of $\alpha$, leading to a continuation
of the 'branch I' of Sect. \ref{Sec4.2} for $\alpha < 0$. Indeed, a negative value of $\alpha$ is a ``wrong sign'' 
choice since it yields a repulsive linear potential of localized solutions in the spherically-symmetric 
case \cite{BrihayeVerbinSph}. 
However, the attractive contribution from the negative cosmological constant is dominant. 
Therefore we do not exclude this possibility. 

The corresponding data is shown on Fig. \ref{critical0-1}.
For fixed $\kappa$, we observe that the parameter $\delta$ decreases monotonically with $\alpha$
and reaches $\delta = 0$ at a minimum value $\alpha = \alpha_m(\kappa)$. For example, we find
$\alpha_m(12) \sim -1.25$ and $\alpha_m(20) \sim -12.0$). For $\alpha <  \alpha_m$,
 Eq. (\ref{melvin}) develops singular point corresponding to $\delta(r) = 0$ and cannot be integrated
 numerically. The difference of the geometric functions $\delta(r)$ and $R(r)$ for the two signs of $\alpha$
 can be appreciated from Fig. \ref{melvin3}.

\begin{figure}[!t]
\centering
\leavevmode\epsfxsize=10.0cm
\epsfbox{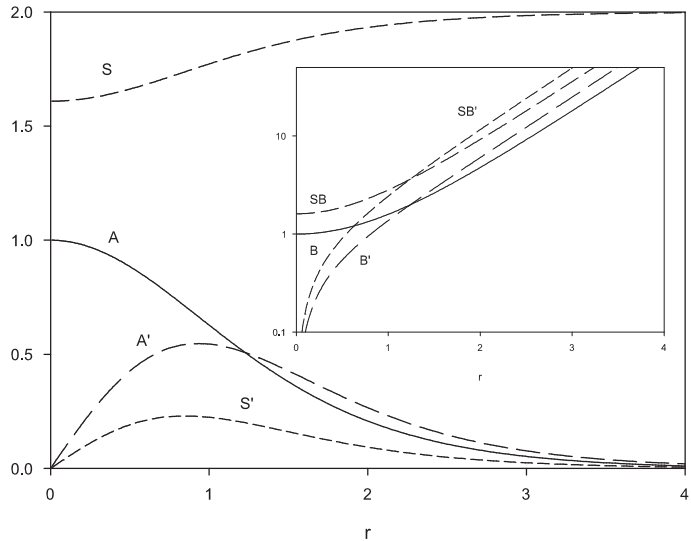}\\
\caption{\label{STMelvin} \small{Profiles of a Melvin-like solution in scalar-tensor  CG. $\alpha =1$,
 $\nu=1$ and $\kappa = 24$.}}
\end{figure}

\subsection{Solutions of the Scalar-Tensor Theory} \label{STEM}

In order to get the Melvin-like solutions in the scalar-tensor extension of the theory, one may start with the general
field equations of the Higgs model of sec. \ref{SecAbHiggs} -- Eqs. (\ref{ScFEqsCylSymm}), (\ref{VecFEqsCylSymm}) and 
(\ref{GravFieldEqCYL}). We replace $\lambda$ by $\alpha\nu$ and $\sqrt{\alpha}f(r)$ by the real (uncharged) scalar field $S(r)$. 
Consequently, we cross out all the interaction terms containing contributions of the form $(m-A)f/L$. 
We solve the system numerically and find that the solutions 
are quite similar to those of the purely tensorial one. Fig. \ref{STMelvin} presents the gauge potential, 
the scalar field and the metric components as well as those of the ''physical metric'', $S(r)B(r)$ and $S(r)B'(r)$. 
Comparison with Figs \ref{melvin1_a} 
and \ref{melvin1_b}  shows that the additional scalar field does not change much the functions $A(r)$ and $B(r)$.

\section{Conformal Strings} \label{SecStrings}
\setcounter{equation}{0}

We finally considered the equations of the Abelian Higgs model coupled to Weyl gravity. Cosmic strings \cite{VilSh} 
are a typical outcome in any field theory which describes matter in the very early universe, thus serving very 
well the purpose of testing the implications of CG.  

Relativistic magnetic flux tubes in flat space were first obtained by Nielsen and Olesen \cite{nielsen_olesen} and 
their coupling to Einstein gravity was shown to have important cosmological consequences \cite{VilSh}. 
A complete classification of the solutions of the self-gravitating Abelian Higgs model was performed \cite{verbin} 
and two branches of solutions were shown to exist: the ``string branch'' consists of solutions which tend asymptotically to 
Minkowski space-time with an angular deficit, and a ``Melvin branch'' whose geometry is
asymptotically similar to that of the Melvin solution. 
Gravitating cosmic strings in the presence of  a cosmological constant, were studied in \cite{bbh}. 

Cosmic strings in the Abelian Higgs model with a conformal coupling to gravity have been also considered before 
(see e.g. \cite{Verbin1999} and references therein), but only within the context of GR.

Now we turn to the analogous solutions within CG, namely solving the full system of the three equations 
(\ref{ScFEqsCylSymm}), (\ref{VecFEqsCylSymm}) and (\ref{GravFieldEqCYL}).
 
Since we are interested in localized solutions, the components of $T^{\mu}_{\nu}$ should vanish asymptotically. 
This, together with the boundary conditions $f(r)\rightarrow v$ and  $A(r)\rightarrow m$ as $r\rightarrow \infty$ 
fix the asymptotic Ricci tensor to be given by (\ref{AsymptAdS}) with $\kappa = 6\lambda v^2$.
Note that unlike the ordinary cosmic strings, here the asymptotic value 
$f(\infty)=v$ does not originate from the Lagrangian, but is a free parameter which characterizes the various 
solutions of the given system.

The system of three equations has to be solved with the boundary conditions
\be \label{CSBC}
f(0)=0,\ \ \ f(\infty)=v,\ \ \ A(0)=0,\ \ \ A(\infty)=m 
\ee
for the matter functions 
and (\ref{bc}) for the metric function $B(r)$. The equations lead to a several 
possible asymptotic forms for the solutions, allowing a few kinds of exponential corrections. 

 For $|\alpha| < 2$,  the form  below appears to be the one which best  fits  the
 numerical solutions obeying the boundary conditions (\ref{CSBC}). 
 For the metric function $B(r)$, we have the same form as before~: 
\be
   \log(B(r\to \infty)) = \gamma_0 + \gamma_1 r + \gamma_2  e^{-\beta r} \ \ ,  \ \  \gamma_1 = \sqrt{\frac{\kappa}{12}} \ \ ,
\ee
where $\gamma_0$, $\gamma_2$ and $\beta$ are constants. 
The parameter $\beta$ encoding the decay rate of the first correction now depends on $\alpha$.
For $|\alpha| < 2$ we find a linear dependance
\be
\label{alpha-beta}
 \beta = \gamma_1 + p\ \alpha \ ,
\ee
where the numerics shows $p \sim 0.18$ . For the matter fields, the asymptotic form is~:
\be
\label{cs_asymp}
 f(r\to \infty) = 
 v\left(1 + F_0 e^{-4 \gamma_1 r} - \frac{\gamma_2 \beta(3\gamma_1- \beta)}{3\gamma_1(\beta+\gamma_1)}e^{-\beta r}\right) 
 \ \ ,   \ \ A(r\to \infty) = A_0 e^{-\gamma_1 s r }  
\ee
with $\gamma_1 = \sqrt{\lambda v^2/2}$ , $s = (1+ \sqrt{1+8/\lambda})/2$ and where $F_0,A_0$ are constants.
We limit our analysis to the case $\lambda = 1$, so that $s = 2$.
The third term appearing in $f$ in (\ref{cs_asymp}) is specific to CG because 
it is related to a source term
appearing due to the  coupling between the scalar field and the Ricci scalar $R$;
it dominates the standard homogeneous piece (the term proportional to $F_0$) since $|\beta| < 4 \gamma_1$. 

The asymptotic form (\ref{cs_asymp}) leads to several qualitative differences with respect to the Melvin case
treated in the previous section.
The most significant difference resides in the fact that the energy momentum tensor decays according to
$T_0^0 \propto \exp{(-\beta r)}$. This implies in particular (see (\ref{alpha-beta}))
\be
\sqrt{-g} \ T_0^0  \ \propto \ e^{(2 \gamma_1 -p\ \alpha )\ r}  \ .   
\ee 
Within the set of parameters which we were able to explore, the exponent is positive and as a consequence, 
the inertial mass (per unit length) of these cosmic string solutions in CG is divergent. 
The presence of the conformal coupling and constant asymptotic curvature (``cosmological constant'')
 has therefore important consequences on the physical 
characteristics  of the cosmic strings; similar phenomena were observed in the context of spherically-symmetric
topological defects, namely for the magnetic monopole \cite{bhrs} and the corresponding CG case \cite{BrihayeVerbinYM}. 

Another different feature of cosmic strings  with respect to the pure magnetic (Melvin) solutions
appears in the asymptotic behavior of the function $\delta(r)$~:
\be
\delta(r \to  \infty) = \beta^2 \gamma_2 \ e^{(\gamma_0- p\ \alpha) r } \ .
\ee 
This combination of the metric function is not any longer constant for $r\to \infty$. This is observed
and confirmed  by the numerical solutions.


\begin{figure}[!b]
\centering
\leavevmode\epsfxsize=10.0cm
\epsfbox{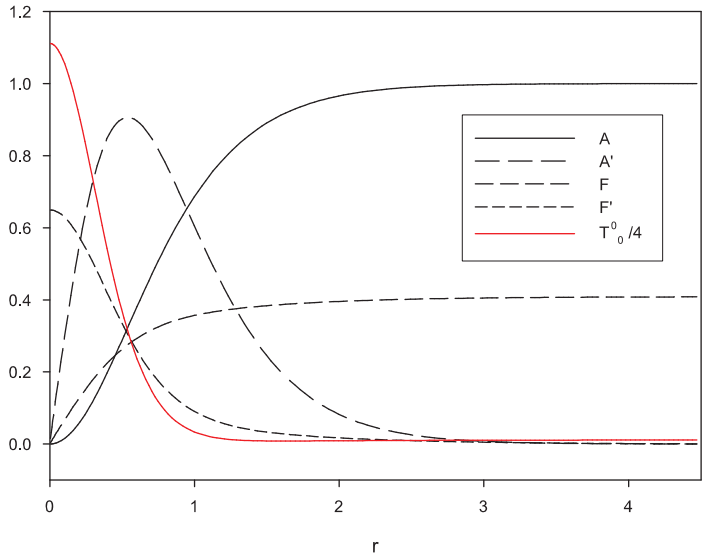}\\
\caption{\label{string3_0} \small{Flux tube solution: Profiles of the matter functions $A,\ A',\ f, \ f'$ and of the energy 
density $T_0^0$ in the case $\alpha=0, \lambda=1$ and $\kappa = 20$.}}
\end{figure}

\begin{figure}[!b]
\centering
\leavevmode\epsfxsize=10.0cm
 \includegraphics[height=.30\textheight,width=.48\textwidth]{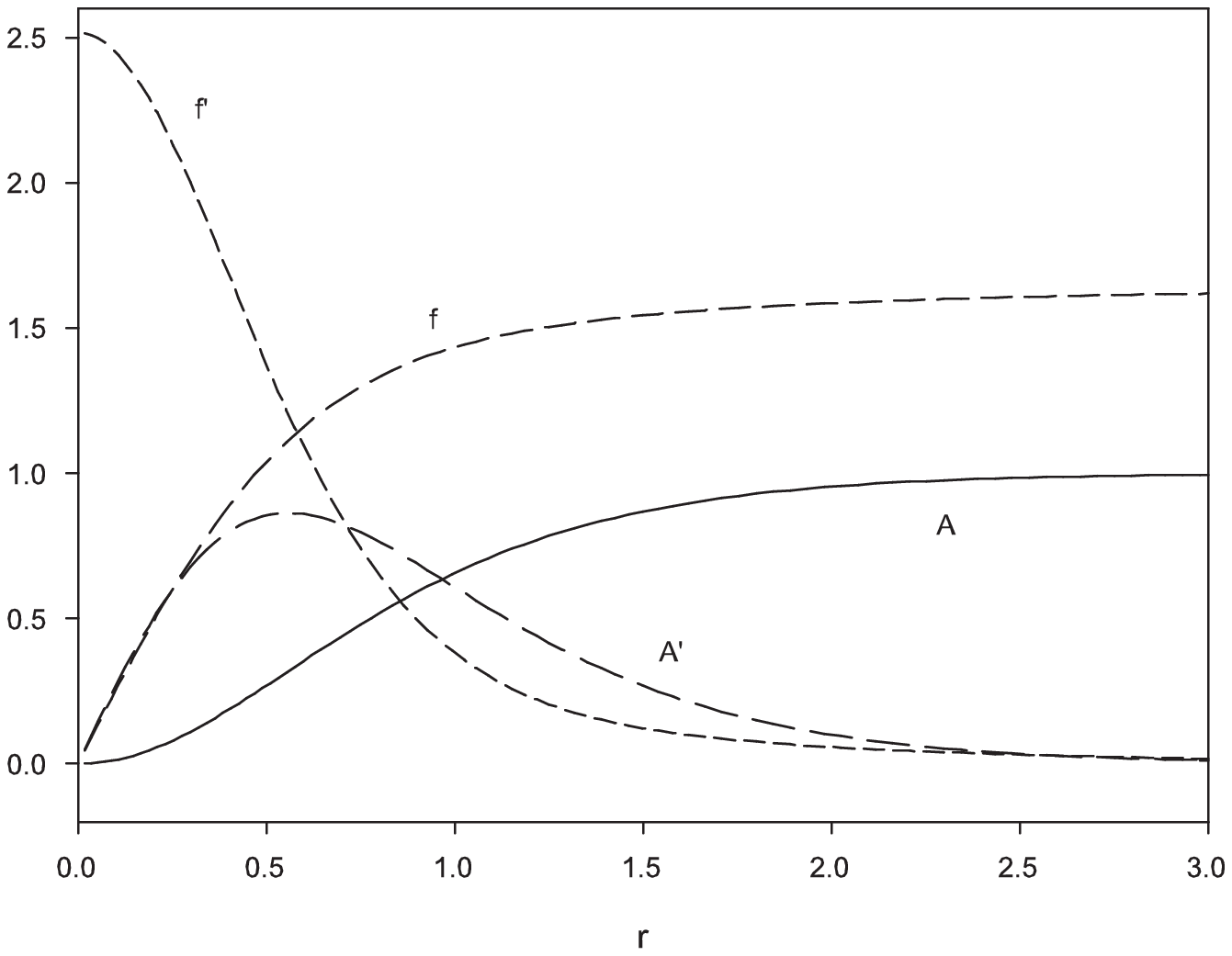}  
 \includegraphics[height=.30\textheight,width=.48\textwidth]{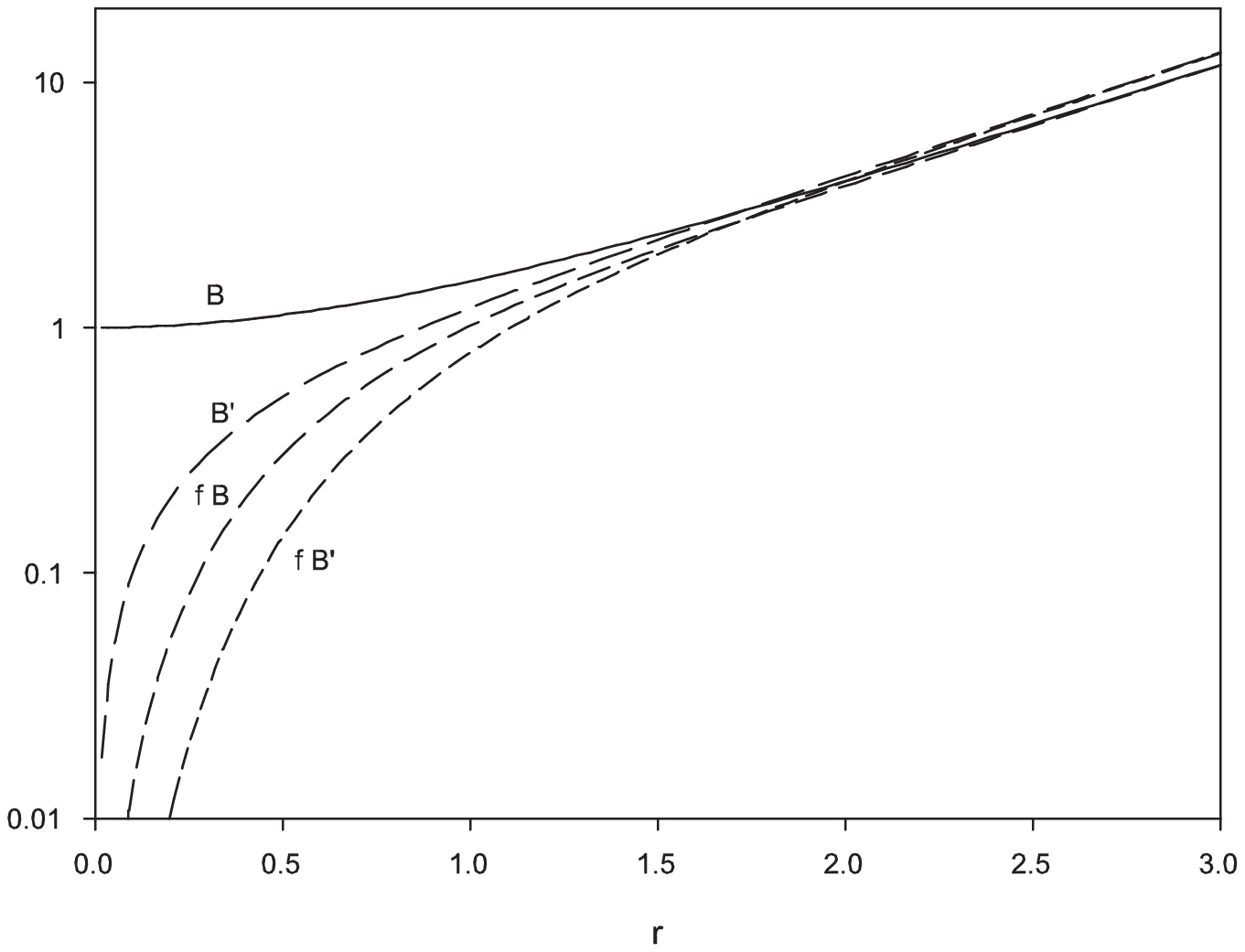}\\
 (a)\hskip 7.5cm (b)\\
 \caption{\label{SGFluxTube} \small{Profiles of a string-like solution for $\alpha=1$, $\lambda=1$ and $\kappa=16$:
(a) The matter functions $A,\ A',\ f, \ f'$;
(b) The metric components $B$, $L=B'$ and those of the physical metric $fB$ and $fB'$.}}
\end{figure}  

We first integrated the equations of conformal string numerically in the case $\alpha=0$ and
 obtained a family of cosmic string  solutions in the background of the vacuum  space-time
characterized by the parameter $\kappa$.
The profiles of the different matter functions of the string embedded in the
background of the vacuum solution with $\kappa=20$ are presented
in Fig. \ref{string3_0} (the energy density is also supplemented).
The profiles of the matter fields are qualitatively very similar to those of the flat case \cite{nielsen_olesen}
or in the gravitating case \cite{verbin}. Considering several values of $\kappa$ leads to similar plots.

Finally, we studied the self-gravitating cosmic strings by integrating the field equations for $\alpha \neq 0$
and found a pattern qualitatively similar to the one obtained for pure electromagnetic (Melvin) solutions. The 
matter fields and metric components of a typical solution are depicted in Fig. \ref{SGFluxTube}. In this case we let
the Higgs field play the role of the scalar $S$, so we stay with the more economic model where the Higgs field itself
is related to mass generation already at the classical level for point particles.

\begin{figure}[!t]
\centering
\leavevmode\epsfxsize=10.0cm
 \includegraphics[height=.30\textheight,width=.48\textwidth]{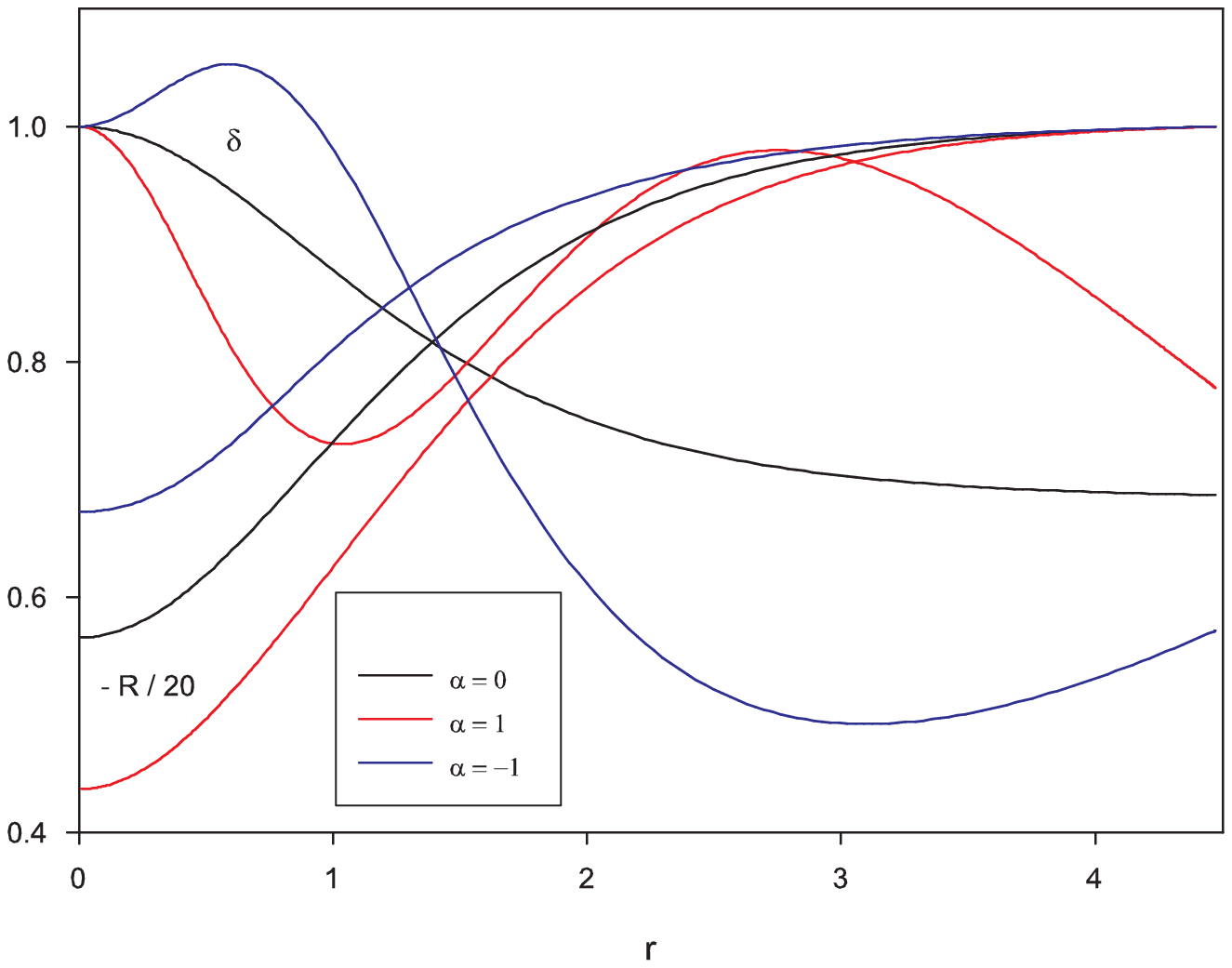}  
 \includegraphics[height=.30\textheight,width=.48\textwidth]{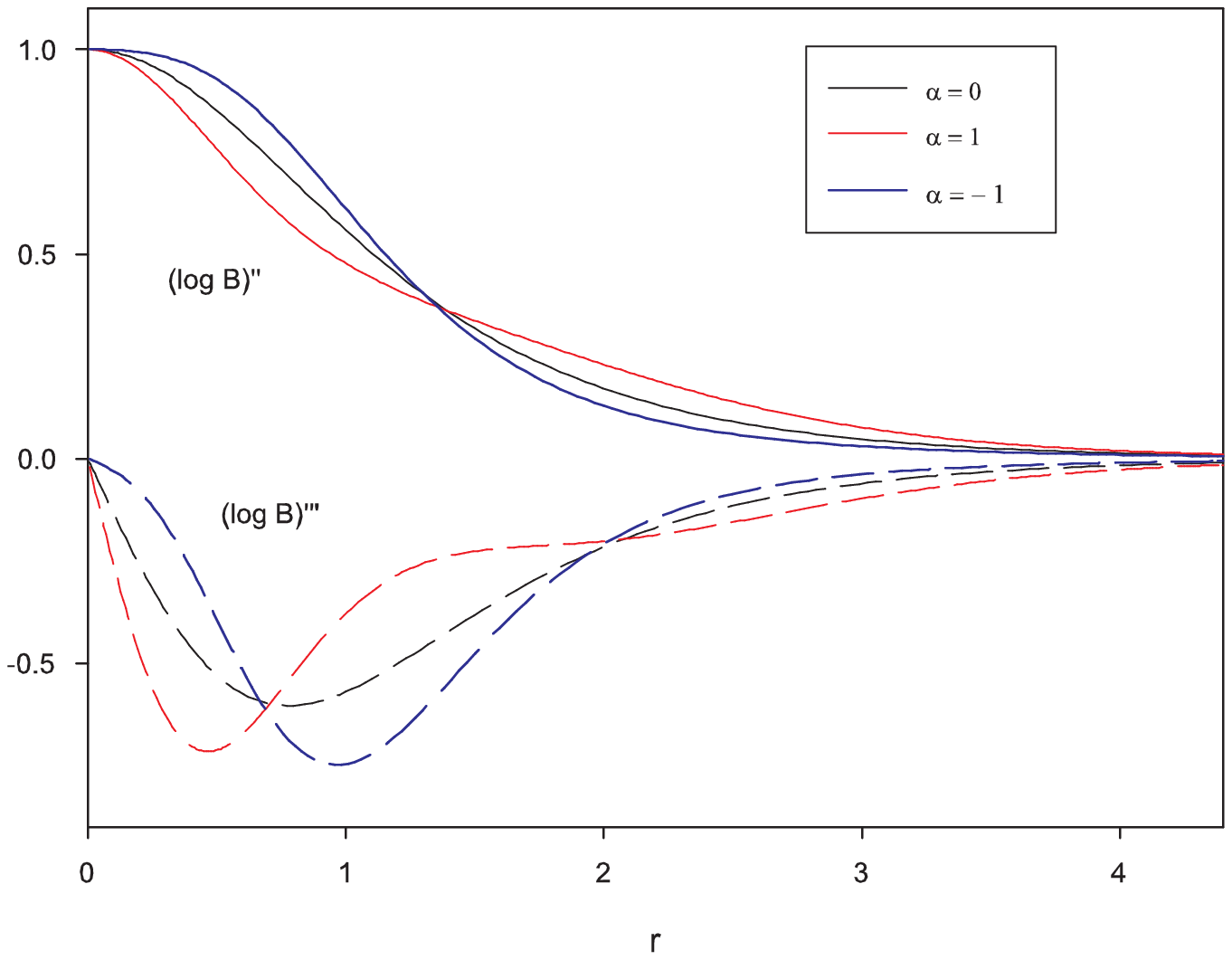}\\
 (a)\hskip 7.5cm (b)\\
 \caption{\label{string3bc} \small{String-like solution for $\lambda=1$ and $\kappa=20$ and
  $\alpha=0$ (black), $\alpha=1$ (red) and $\alpha=-1$ (blue):
(a) Profiles of $R(r)$ and $\delta(r)$. Note that the $\delta(r)$ curve for $\alpha=0$ can be fit between the 
two curves of Fig. \ref{melvin1_a}. ;
(b) Details of derivatives of $B(r)$.}}
\end{figure}

Although the profiles of the matter functions deviate a little from the $\alpha=0$ case,
 the deformation of the metric function $B(r)$ and some quantities characterizing the geometry are more significant.
The quantities $R(r), \delta(r)$ are compared in Fig. \ref{string3bc}a for $\alpha=0$
and $\alpha = \pm 1$. The derivatives $B'',B'''$ also develop some structure close to the symmetry axis $r=0$;
this is illustrated in Fig. \ref{string3bc}b.   

In this case also, the gravitating cosmic string solutions exist only up to a maximal value of $\alpha$.
E.g. we find solutions for $\alpha < 1.785$ (resp. $\alpha < 1.2$) for $\kappa=20$ (resp. $\kappa=16$).
We strongly suspect that, like in the Melvin case, a second branch of solutions exist but
it was  not attempted to construct it.
The determination of the domain of existence of the solutions in the $\kappa-\alpha$ plane was
also left for further investigation. 

\section{Conclusions} \label{concl}

In this paper we have investigated cylindrically-symmetric solutions in Conformal Gravity. Examining the solutions 
both by analytical and numerical methods we were able to construct several
types of solutions which to our knowledge were unknown so far: soliton-like regular vacuum 
solutions, spaces around line sources, the CG analogs of the Melvin solution and conformal self-gravitating flux tubes, 
i.e. the CG analogs of the cosmic strings

The analysis was based on the Abelian Higgs model coupled to CG which was studied first in several special cases and finally
for cosmic strings. The symmetry breaking mechanism which is required for flux 
tube solutions has in the present case a gravitational origin, unlike the ordinary case, since conformal symmetry 
does not allow the usual negative mass term into the Lagrangian. 

The Higgs model was also used as a starting point of extending CG to be a scalar-tensor theory which allows  a
consistent coupling of point particles.

CG is a rich and interesting theory which may supply answers to some of the most annoying problems 
like those of the cosmological constant, the dark matter and the dark energy. It is therefore natural and required to
investigate the theory in other domains and search for further implications of the theory like in the area of
topological defects which are assumed to play an important role on structure formation at the early universe.

As a first step we studied the vacuum solutions of CG which are interesting on their own right and also
as they give a first indication about the asymptotic behavior of localized self-gravitating structures. We found that the 
cylindrically-symmetric vacuum open solutions of CG have an asymptotical behavior which is similar to that of the AdS soliton. 
Actually there are two kinds of solutions: one with an everywhere constant negative (``AdS-like'') Ricci curvature 
(the same as the AdS soliton) and the second, which is a new kind special to CG, has a Ricci scalar which approaches  a 
constant only asymptotically. 
There exist also two kinds of flat or asymptotically flat vacuum solutions which we have not studied extensively, since
they are not compatible with the symmetry breaking mechanism which produces in this context the flux tube solutions.

The second step was to study magnetic solutions in the purely tensorial CG as well as in its scalar-tensor 
extension. We have found  Melvin-like solutions that have a finite energy per unit length. The solutions
separate into two branches and in each of them  CG cannot support magnetic fluxes  with too intense magnetic fields in the core.

The last step was centered about self-gravitating flux tube solutions in CG. Here the system contains from the start a
scalar field which may be naturally exploited for consistent coupling of point particles, so we did not extend the 
model by an additional scalar field. We were able to solve numerically the field equations and obtained 2 important results:
(i) As in the purely magnetic solutions, there is an upper bound on the magnetic field in the core above which no solutions 
exist.
(ii) Unlike the purely magnetic case, the Ricci scalar (``cosmological constant'') influences drastically the asymptotic 
decay of the scalar fields and makes it impossible to have a finite inertial mass.

\end{document}